\documentclass[12pt]{article}
\usepackage[utf8]{inputenc}
\usepackage{mathptmx}
\usepackage{geometry}
\usepackage{authblk}
\usepackage{amssymb}
\usepackage{amsmath}
\usepackage{amsthm}
\allowdisplaybreaks[4]
\usepackage{slashed}
\usepackage{graphicx,color,epsfig}
\usepackage{multirow}
\usepackage{cite}
\usepackage[colorlinks=true,linkcolor=red,citecolor=blue]{hyperref}
\begin{document}
\renewcommand{\arraystretch}{0.45}
\newcommand{\beq}{\begin{eqnarray}}
\newcommand{\eeq}{\end{eqnarray}}
\newcommand{\non}{\nonumber\\ }
\newcommand{\acp}{ {\cal A}_{CP} }
\newcommand{\psl}{ p \hspace{-1.8truemm}/ }
\newcommand{\nsl}{ n \hspace{-2.2truemm}/ }
\newcommand{\vsl}{ v \hspace{-2.2truemm}/ }
\newcommand{\epsl}{\epsilon \hspace{-1.8truemm}/\,  }
\def\ra{\rangle}
\def\la{\langle}
\def\sl{\!\!\!\!\slash}
\def\ov{\overline}
\newcommand{\tf}{\textbf}
\title{Analysis of CKM-Favored Quasi-Two-Body $B  \to D (R\to) K \pi$ Decays in PQCD Approach}
\author[1]{Zhi-Tian Zou}
\author[1]{Wen-Sheng Fang}
\author[2]{Xin Liu}
\author[1]{Ying Li$\footnote{liying@ytu.edu.cn}$}
\affil[1]{\it \small Department of Physics, Yantai University, Yantai 264005, China}
\affil[2]{\it \small School of Physics and Electronic Engineering, Jiangsu Normal University, Xuzhou 221116, China}
\maketitle
\begin{abstract}
LHCb Collaboration studied the resonant structure of $B_s\to \overline{D}^0K^-\pi^+$ decays using the Dalitz plot analysis technique, based on a data sample corresponding to an integrated luminosity of $3.0{\rm fb}^{-1}$ of $pp$ collision. The $K^-\pi^+$ components have been analyzed in the amplitude model, where the decay amplitude is modeled to be the resonant contributions with respect to the intermediate resonances $K^*(892)$, $K_0^*(1430)$ and $K_2^*(1430)$. Motivated by the experimental results, we investigate the color-favored quasi-two-body $B \to \overline{D}^0K\pi$ decays in the framework of the perturbative QCD (PQCD) approach. We calculate the the branching fractions by introducing the appropriate wave functions of $K\pi$ pair. Our results are in agreement well the available data, and others can be tested in LHCb and Belle-II experiments. Using the narrow-width-approximation, we also extract the branching fractions of the corresponding two-body $B\to \overline D R$ decays, which agree to the previous theoretical calculations and the experimental data within the errors. There are no $CP$ asymmetries in these decays in the standard model, because these decays are all governed  by only the tree operators.
\end{abstract}

\section{Introduction}
Three-body non-leptonic $B$ meson decays constitute a large portion of the branching fraction and therefore attract a lot of attentions for several phenomenological applications, such as deeply testing the standard model (SM), the study of $CP$ violation and the exaction of the CKM angles \cite{Grossman:2002aq}. The branching fractions and $CP$ asymmetries of a large number of channels have been studied extensively by BaBar \cite{BaBar:2009vfr, BaBar:2012iuj, BaBar:2009jov,BaBar:2011vfx, BaBar:2011ktx, BaBar:2008lpx, BaBar:2009vfr}, Belle \cite{Belle:2010wis, Belle:2008til, Belle:2006ljg,Belle:2005rpz} and LHCb \cite{LHCb:2019sus, LHCb:2019jta, LHCb:2017hbp, LHCb:2019vww, LHCb:2016vqn}. Stimulated by the abundant experimental measurements, the theoretical explorations have been carried out in the recent years. Many three-body nonleptonic decays of heavy mesons have been studied in detail in QCD factorization \cite{El-Bennich:2009gqk, Krankl:2015fha, Cheng:2002qu, Cheng:2016shb, Cheng:2014uga, Li:2014oca, Li:2014fla}, PQCD approach~ \cite{Wang:2014ira, Wang:2016rlo, Li:2016tpn, Rui:2017bgg, Zou:2020atb, Zou:2020fax, Zou:2020mul, Zou:2020ool, Yang:2021zcx, Liu:2021sdw} and other theoretical methods \cite{Zhang:2013oqa, El-Bennich:2006rcn, Cheng:2019tgh, Hu:2022eql}.

Different from the two-body $B$ decays where the momenta of final states are fixed, The kinematics of the three-body decay is completely determined by two of the three kinematic invariants $s_{12}$, $s_{13}$ and $s_{23}$, with the definition $s_{ij}=(P_i + P_j)^2/M_B^2$. The physical kinematical region in the plane of two invariants (the Dalitz plot) is given by a triangle. The Dalitz plot analysis has been proved to be a powerful tool for studies of multi-body decays, and has been applied widely in the experimental analysis. The analysis of three-body $B$ decays using this technique enables one to study the properties of various resonances. Indeed, most of the quasi-two-body $B$ decays are extracted from the Dalitz-plot analysis of three-body ones. In addition, the Dalitz plot analysis of three-body $B$ decays provides a nice methodology for extracting information on the unitarity triangle in SM. For example, the Dalitz analysis together with the isospin symmetry allows one to extract the angle $\alpha$ from the $B\to \pi\pi\pi$ decays with the vector resonance $\rho$ \cite{Snyder:1993mx}.

On the theoretical side, three-body decays of heavy $B$ mesons are rather more complicated as they receive resonant and nonresonant contributions and involve 3-body matrix elements. The interference between nonresonant and resonant amplitudes makes it difficult to disentangle these two distinct contributions clearly. The experimental measurements show that some three-body $B$ decays are dominated by intermediate vector, scalar and tensor resonances, namely, they proceed via quasi-two-body decays containing a resonance state and a bachelor. While for some others, the nonresonant contributions govern the decay amplitudes \cite{Krankl:2015fha, Cheng:2007si}. Summarizing the experimental measurements, we find that there is a large nonresonant fraction in the penguin-dominant modes. For example, the nonresonant fraction reaches about $90\%$ in $B\to KKK$ decays, $17-40\%$ in $B\to K\pi\pi$ decays, and only $14\%$ in $B\to \pi\pi\pi$ decays \cite{Cheng:2008vy}. However, there is not any reliable theoretical description of the amplitudes of $B$ meson three-body decays so far, and most studies is are still model dependent. The isobar model \cite{Herndon:1973yn} and the K-matrix formalism \cite{Chung:1995dx} are popularly adopted in the experimental analysis, especially the former \cite{LHCb:2019sus}. In these models, the resonant amplitudes are modeled and the non-resonant contributions are often described by an empirical distribution in order to reproduce the whole phase space \cite{Belle:2004drb}.

Besides a large number of charmless three-body $B$ decays, a lot of charmed three-body decays have also been measured in $B$ factories \cite{BaBar:2009pnd} and LHCb \cite{LHCb:2014ioa, LHCb:2018oeb}. For example, $B_s\to \overline{D}^0K^-\pi^+$ have been detailed analyzed in LHCb in ref.\cite{LHCb:2014ioa}, where the resonant structures of the $\overline{D}^0K^-$ and $K^-\pi^+$ components have been studied using the Dalitz plot technique. For the $\overline{D}^0K^-$ component, a structure is found at $m_{\overline{D}^0K^-}\approx 2.86$ GeV, which is viewed as an admixture of spin-1 and spin-3 resonances. For the $K^-\pi^+$ component, $K^*$ resonances and the corresponding fit fractions have been well measured and the branching fractions of the corresponding quasi-two-body decays were also reported, based on the fit fraction and the branching fraction of the $B_s\to \overline{D}^0K^-\pi^+$ from previous LHCb measurement \cite{LHCb:2013svv}. Motivated by this, we shall investigate the CKM-favored $B_{(s)}\to \overline{D}K \pi$ decays in PQCD approach, and focus on the $S$, $P$ and $D$-wave resonant contributions with respect to the $K\pi$ component.

We all know that the resonant three-body decays (quasi-two-body decays) correspond to the edge of the Dalitz plot, where the two particles move collinearly with large energy and the bachelor particle recoils back. The meson-pair and the rest one move fast and back-to-back in the $B$ meson rest frame so that the interaction between the meson-pair and the bachelor meson is highly suppressed. If we regard the meson-pair as a whole, the quasi-two-body decay is very similar to a two-body decay and the factorization formula for two-body decay can be applied safely. In practice, we also need to introduce a new wave function to describe the meson-pair, including the angular momenta. The soft interactions between two mesons are also absorbed in this nonperturbative wave function.

In PQCD approach, the amplitude of quasi-two-body $B$ decay can be factorized into different parts according to the ``characteristic" scales. As we know, the physics above the mass of $W$ boson $m_{W}$ is weak interaction and can be calculated perturbatively. The physics between $m_{W}$ scale and $b$ quark mass $m_b$ scale can be contained in the Wilson coefficient $C(\mu)$, which can be calculated within the renormalization group equation and the Wilson coefficient at the scale $m_{W}$. The physics between $m_b$ scale and the factorization scale $\Lambda_h$ is dominated by exchanging one hard gluon and can be calculated perturbatively. This part is also called hard kernel $H$. The physics below  $\Lambda_h$ scale is soft and nonperturbative, which can be described by the universal wave functions. Therefore, the decay amplitude  of $B \to D (R\to) K \pi$ decay in PQCD can be decomposed into the convolution as \cite{Keum:2000ms}
\begin{eqnarray}
\mathcal{A}\sim\int dx_idb_i\Big[C(t)\otimes{H}(x_i,b_i,t)\otimes\Phi_B(x_1,b_1)\otimes\Phi_D(x_2,b_2)\otimes\Phi_{K\pi}(x_3,b_3)\otimes e^{-S(t)}\Big], \label{fq}
\end{eqnarray}
where $x_i$ are the momentum fraction of the quarks in the initial and final mesons, and $b_i$ are the conjugate variables of the intrinsic transverse momentum $k_{Ti}$ of the quarks. $\Phi_B$, $\Phi_D$ are the wave functions of the $B$ meson and the $\overline{D}$ meson, respectively. $\Phi_{K\pi}$ is the new introduced two-meson wave function. The exponential term $e^{-S(t)}$ is the Sudakov form factor obtained from the resummation of the double logarithms arising from the retained intrinsic transverse momentum $K_T$ of the inner quarks \cite{Keum:2000ph,Lu:2000em,Li:2003yj}.

The layout of the present paper is as follows. In Sec.\ref{sec:function},  we firstly show the Hamiltonian governing those quasi-two-body $B \rightarrow \overline{D}K\pi$ decays considered in this paper. The wave functions of $K-\pi$ pair are also discussed. In Sec.\ref{sec:result} we address the numerical results of the branching fractions of those quasi-two-body decays. In this section based on the obtained branching fractions we will also probe the corresponding two-body $B \to \overline{D} R$ ($R$ denoting the intermediate resonance) decays and compare them with the previous results and the experimental data. Finally, we conclude this work in Sec.\ref{sec:summary}
\section{The Decay Formalism and Wave Function}\label{sec:function}
For the quasi-two-body decays, the Dalitz plot analysis help us to interpret the decay amplitude in isobar model conveniently, which is popularly used to describe the complex amplitude of three-body decays by the experimentalists. In this model the total decay amplitude is represented by a coherent sum of amplitudes from $N$ individual decay channels with different resonances,
\begin{eqnarray}
\mathcal{A}=\sum_{i=1}^{N}a_i\mathcal{A}_i,
\end{eqnarray}
where $a_i$ is the complex coefficient showing the relevant magnitude and the phase of different decay channels. The phase is a new source of the $CP$ asymmetry, namely, the $CP$ asymmetry caused by the interference among the different channels. $\mathcal{A}_i$ is the amplitude of the quasi-two-body decay  associated with the certain resonance, which can be perturbatively calculated in PQCD.

The weak Hamiltonian $\mathcal{H}_{eff}$ of $\bar b\to \bar c{u}\bar q$ governing the $B_{(s)}\to\overline{D}K\pi$ decays can be expressed as~\cite{Buchalla:1995vs}
\begin{eqnarray} \label{hamiltionian}
\mathcal{H}_{eff}=\frac{G_F}{\sqrt{2}}\Big\{V_{uq}^*V_{cb}(C_1O_1+C_2O_2)\Big\}\left |_{q=d, s}\right.,
\end{eqnarray}
where $V_{cb}$ and $V_{uq}$ are the CKM matrix elements. The $C_{1}$ and $C_2$ are the so-called Wilson coefficients corresponding to the four-quark operators $O_1$ and $O_2$ respectively, which are all tree operators. The explicit expressions of the operators are written as
\begin{eqnarray}
O_1=(\bar{b}_{\alpha}c_{\beta})_{V-A}(\bar{u}_{\beta}q_{\alpha})_{V-A};\;\;\;O_2=(\bar{b}_{\alpha}c_{\alpha})_{V-A}(\bar{u}_{\beta}q_{\beta})_{V-A}.
\end{eqnarray}
We all know that the direct $CP$ asymmetry is related to the interference between the tree type contribution and that of the penguin operators. Therefore, there are no direct local $CP$ asymmetries in decays induced by the $\bar b\to \bar c{u}\bar q$ transition in SM, because the penguin operators do not involved in these decays.

It can be found from the factorization formula as shown in Eq.(\ref{fq}) that the most important inputs are the wave functions of the initial and final mesons. For the initial $B$ meson and the final $\overline{D}$ meson, their wave functions have been well determined and adopted widely in the studies of two-body $B_{(s)}\to PP,PV,VV$ \cite{Lu:2000em, Yu:2005rh, Ali:2007ff, Zou:2015iwa} and $B_{(s)}\to DP,DV,DS,DT$ \cite{Zou:2009zza, Li:2008ts, Zou:2017yxc, Zou:2017iau, Zou:2016yhb, Zou:2012sy, Zou:2012sx} decays and other decay processes \cite{Zou:2012td, Liu:2013cvx, Wang:2017hxe}, and we will not discuss them any more in this paper. In the quasi-two-body decays, the new introduced parameters are the two-meson wave functions corresponding to different resonances. In current work, according to the LHCb analysis \cite{LHCb:2014ioa}, we will consider the resonant contributions from the scalar resonance $K^*_0(1430)$, the vector resonances $K^*(892)$ and $K^*(1410)$, and the tensor resonance $K^*_2(1430)$ to the $B_{(s)}\to \overline{D} K\pi$ decays. So, three kinds of wave functions will be discussed in the following sections. It should be noted that at present the forms of the two-meson wave functions based on QCD-inspired approach are still absent, and many phenomenological attempts have been performed to determine these forms and to constrain the involved parameters based on the available experimental data.

We firstly discuss the $S$-wave $K\pi$ pair wave function $\Phi_{S,K\pi}$, which can be given as \cite{Li:2019hnt},
\begin{eqnarray}
\Phi_{S,K\pi}=\frac{1}{2\sqrt{N_c}}\left[P\mkern-10.5mu/\phi_S(z,\zeta,\omega)+\omega\phi_S^s(z,\zeta,\omega)+\omega(n\mkern-10.5mu/ v\mkern-10.5mu/-1)\phi_S^t(z,\zeta,\omega)\right],
\end{eqnarray}
with $P$ and $\omega$ denoting the momentum and the invariant mass of the $K\pi$ pair respectively, satisfying $P^2=\omega^2$. The light-like vectors $n=(1,0,\mathbf{0}_T)$ and $v=(0,1,\mathbf{0}_T)$ are the dimensionless vectors. $\phi_S$ and $\phi_S^{s,t}$ are the twist-2 and twist-3 light-cone distribution amplitudes (LCDAs), respectively. The inner parameter $z$ is the momentum fraction of the spectator quark, and $\xi$ is the momentum fraction of the $K$ meson in the $K\pi$ pair. Similar to the LCDA of the meson, the LCDAs of the $K\pi$ pair can also be decomposed into the Gegenbauer polynomials together with the corresponding Gegenbauer moments, which are given as
\begin{eqnarray}
\phi_S(z,\xi,\omega)&=&\frac{6}{2\sqrt{2N_C}}F_S(\omega)z(1-z)\left[\frac{1}{\mu_S}+B_1C_1^{3/2}(1-2z)+B_3C_3^{3/2}(1-2z)\right],\\
\phi_S^s(z,\xi,\omega)&=&\frac{1}{2\sqrt{2N_c}}F_S(\omega),\\
\phi_S^t(z,\xi,\omega)&=&\frac{1}{2\sqrt{2N_C}}F_S(\omega)(1-2z),
\end{eqnarray}
with
\begin{eqnarray}
\mu_S=\frac{\omega}{m_2-m_1},
\end{eqnarray}
where $m_{1,2}$ are the masses of the running current quarks in the resonance $K_0^*(1430)$. $C_{1,3}^{3/2}(t)$ and $B_{1,3}$ are the Gegenbauer polynomials and corresponding Gegenbauer moments respectively. For the twist-3 LCDAs, the asymptotic forms are adopted for simplicity. However, the values of the inner parameters $B_{1,3}$ in twist-2 LCDA have not been determined by the QCD-inspired approach so far. We determined them to be $B_1=-0.4\pm0.2$ and $B_3=0.7\pm0.4$ by phenomenological method together with the available experimental measurements of the $B_s\to \overline{D}^0K^-\pi^+$ decay \cite{LHCb:2014ioa}. In ref.\cite{Zou:2020fax}, we had investigated the $B_s\to K^0(\overline{K}^0)K^{\pm}\pi^{\mp} $ and fitted $B_1=-0.4\pm0.2$ and $B_3=-0.8\pm0.4$. It is obvious that the values of $B_1$ are almost same and the signs of $B_3$ are different.  We also note that, in previous studies of the charmless $B_s\to K^0(\overline{K}^0)K^{\pm}\pi^{\mp} $ decays the terms proportional to the fraction $(\frac{m_i}{M_{B}})^2$ have been neglected since the mass of the final $K$ or $\pi$ is small enough to be omitted, in comparison with the mass $M_{B}$ of $B$ meson. In this work, when we consider the charmed $B$ meson decays with massive $\overline{D}$ meson involved in final states, the terms proportional to the fraction $(\frac{M_D}{M_B})^2$ can not be neglected any more, which will change the behaviour of the propagator of the inner quarks. Compared with $B_1$, $B_3$ has a large fluctuation in the fitting procedure. As a result, the new fitted $B_{3}$ differs from the previous result, which is also confirmed in ref.\cite{Li:2021cnd}. In ref.~\cite{Wang:2020saq}, these parameters were also taken as the same as those in wave function of the $K_0^*(1430)$ meson with large uncertainties considered.

Unlike the distribution amplitudes of meson,  the time-like form factor $F_S(\omega)$ is involved. In particular, this form factor is  parameterized by the relativistic Breit-Wigner (RBW) model which is adopted extensively in experimental analysis, and has been proved to be a valid model for describing the narrow resonance that can be well separated from the other resonant and non-resonant contributions.  However, RBW model fails to describe the $F_S(\omega)$ associated with the $K_0^*(1430)$ resonance, because  $K_0^*(1430)$ resonance interferes strongly with a slowly varying non-resonant contribution \cite{Meadows:2007jm}. In order to overcome this deficiency, the so-called LASS line shape \cite{Aston:1987ir, Back:2017zqt} is developed to describe the combined contributions including the resonant and non-resonant parts, which explicit expression  is given as
\begin{eqnarray}
F_S(\omega)=\frac{\omega}{\mid p_1\mid(\cot\delta-i)}+e^{2i\delta}\frac{m_0\Gamma_0\frac{m_0}{\mid p_0\mid}}{m_0^2-\omega^2-im_0\Gamma_0\frac{\mid p_1\mid}{\omega}\frac{m_0}{p_0}},
\label{lass}
\end{eqnarray}
where the phase $\delta$ is determined by $\cot\delta=\frac{1}{a\mid p_1\mid}+\frac{r\mid p_1\mid}{2}$ with the scattering length $a=1.95\pm0.09~{\rm GeV}^{-1}$ and the effective range $r=1.76\pm0.36~{\rm GeV}^{-1}$ \cite{112}. The $|p_1|$ is the magnitude of the momentum of one of daughter of the resonance $K_0^*(1430)$ in the center-of-mass frame of the $K\pi$ pair and the $|p_0|$ is the value of the $|p_1|$ when $\omega=m_0$. Finally, $m_0$ and $\Gamma_0$ are the pole mass and width of the resonance $K_0^*(1430)$. From Eq.(\ref{lass}), one can easily find that the first term represents the non-resonant contribution, while the second term is the resonant one. Usually, a cutoff at $\omega=1.7~{\rm GeV}$ is suggested for the nonresonant contributions, which implies that there is only resonant contribution when the invariant mass is larger than $1.7~\rm GeV$. In this paper we will also adopt the LASS model to evaluate the $S$-wave contributions to the $B_{(s)}\to \overline{D} K\pi$ decays, which has also been adopted by LHCb.

Phenomenologically, the wave function of the $P$-wave $K\pi$ pair can be modeled from the wave function of the vector meson. Due to the law of conservation of angular momentum, only the longitudinal polarization component contributes to the decay amplitude of $B  \to \overline D (R\to) K \pi$, so we only keep the longitudinal wave function
\begin{eqnarray}
\Phi_{P,K\pi}=\frac{1}{\sqrt{2N_C}}\left[P\mkern-10.5mu/\phi_P(z,\xi,\omega)+\omega\phi_P^s(z,\xi,\omega)
+\frac{P\mkern-10.5mu/_1P\mkern-10.5mu/_2-P\mkern-10.5mu/_2P\mkern-10.5mu/_1}{\omega(2\xi-1)}\phi_P^t(z,\xi,\omega)\right],
\label{pwave}
\end{eqnarray}
where the expressions of twist-2 and twist-3 LCDAs are
\begin{eqnarray}
\phi_P(z,\xi,\omega)&=&\frac{3F_P^{\parallel}(\omega)}{\sqrt{2N_C}}z(1-z)\left[1+a_1C_1^{3/2}(t)+a_2C_2^{3/2}(t)\right](2\xi-1),\\
\phi_P^s(z,\xi,\omega)&=&\frac{3F_P^{\perp}(\omega)}{2\sqrt{2N_C}}t(2\xi-1),\\
\phi_P^t(z,\xi,\omega)&=&\frac{3F_P^{\perp}(\omega)}{2\sqrt{2N_C}}t^2(2\xi-1),
\end{eqnarray}
with $t=1-2z$. For the twist-3 LCDAs $\phi_P^{s,t}$, we also take the asymptotic forms. The Gegenbauer moments $a_{1,2}$ in the twist-2 LCDA are taken as $a_1=-0.4$ and $a_2=0.46$ ~\cite{Li:2021cnd}.

For the time-like form factor $F_P^{\parallel}$ corresponding to the resonances $K^*(892)$ and $K^*(1410)$, it can be well modeled by the RBW model, and be given as \cite{Back:2017zqt}
\begin{eqnarray}
F_{\rm RBW}(\omega)=\frac{m_0^2}{m_0^2-\omega^2-im_0\Gamma(\omega)},
\end{eqnarray}
with the nominal mass $m_0$ of the resonance. The mass-dependent width $\Gamma(\omega)$ corresponding to a resonance with spin $L$ can be expressed as~\cite{BaBar:2012iuj}
\begin{eqnarray}
\Gamma(\omega)=\Gamma_0\left(\frac{|p_1|}{|p_0|}\right)^{2L+1}\left(\frac{m_0}{\omega}\right)X_L^2(r|p_1|),
\label{rbw}
\end{eqnarray}
with the same definitions as the eq.(\ref{lass}) for the $\Gamma_0$, $|p_1|$ and $|p_0|$. The $X_L$ is the Blatt-Weillkopf barrier factor \cite{Blatt:1952ije}, which depends on the angular momentum of the meson pair, with the expression as
\begin{eqnarray}
&&L=0,X_0(a)=1,\\
&&L=1,X_1(a)=\sqrt{\frac{1+a_0^2}{1+a^2}},\\
&&L=2,X_2(a)=\sqrt{\frac{a_0^4+3a_0^2+9}{a^4+3a^2+9}},
\end{eqnarray}
$a_0$ being the value of the $a$ at the pole mass of the resonance. The inner parameter $r$ in eq.(\ref{rbw}) reflects the effective barrier of the resonance, and is usually taken to be $r=4~{\rm GeV}^{-1}\approx 0.8$ fm \cite{BaBar:2005qms} for each resonance, because it dose not affect the numerical results remarkably. Currently, $X_1(a)$ is used for the $P$-wave longitudinal time-like form factor, that is
\begin{eqnarray}
F_P^{\parallel}(\omega)=F_{\rm RBW}(\omega)\mid_{L=1}.
\end{eqnarray}
For the transverse time-like form factor, $F_P^{\perp}(\omega)$ can be obtained using the relation \cite{Wang:2016rlo}
\begin{eqnarray}
\frac{F_P^{\perp}(\omega)}{F_P^{\parallel}(\omega)}\approx\frac{f_V^{\perp}}{f_V},
\label{prell}
\end{eqnarray}
with the decay constants $f_V^{(\perp)}$ of the vector resonance.

Now, we turn to discuss the wave function of the $D$-wave $K\pi$ pair. Because the initial $B$ meson and final state $D$ meson are all pseudoscalars, the helicity $\lambda=\pm2$ components vanish due to the conservation of the angular momentum. Therefore, the behavior of the $D$-wave $K\pi$ pair are very similar to the $P$-wave one \cite{Cheng:2010hn}. The only differences are the LCDAs $\phi_D$, $\phi_D^s$, and $\phi_D^t$, whose expressions are
\begin{eqnarray}
\phi_D(z,\xi,\omega)&=&\sqrt{\frac{2}{3}}\frac{6F_D^{\parallel}(\omega)}{2\sqrt{2N_C}}z(1-z)\left[3a_D(2z-1)\right]P_2(2\xi-1),\\
\phi_D^s(z,\xi,\omega)&=&\sqrt{\frac{2}{3}}\frac{-9F_D^{\perp}(\omega)}{4\sqrt{2N_C}}\left[a_D(1-6z+6z^2)\right]P_2(2\xi-1),\\
\phi_D^t(z,\xi,\omega)&=&\sqrt{\frac{2}{3}}\frac{9F_D^{\perp}(\omega)}{4\sqrt{2N_C}}\left[a_D(1-6z+6z^2)(2z-1)\right]P_2(2\xi-1),
\end{eqnarray}
where the Gegenbauer moment $a_D$ is determined to be 0.25 and function $P_2(t)$ is the Legendre polynomial. $F_D^{\parallel}(\omega)$ and $F_D^{\perp}(\omega)$ are the $D$-wave longitudinal and transverse time-like form factors respectively, which can also be well parameterized by RBW model with the spin-2 resonance $K_2^*(1430)$.

\begin{figure}[!ht]
\begin{center}
\includegraphics[scale=0.6]{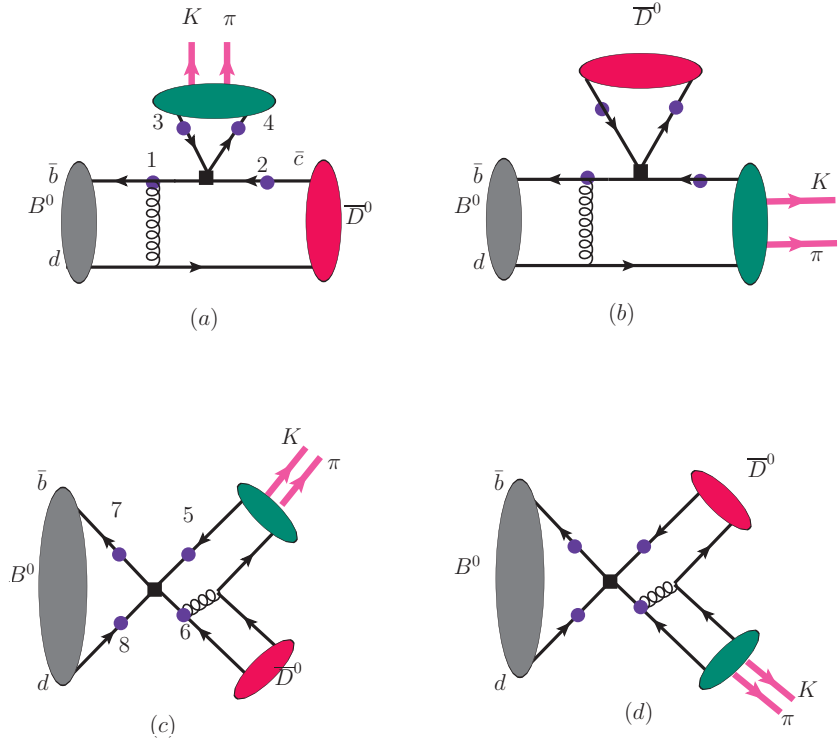}
\vspace{-6cm}
\caption{Typical Feynman diagrams for the quasi-two-body decay $B^0\to  \overline D^0(R\to)K\pi$ in PQCD, where the black squares stand for the weak vertices, large (purple) spots on the quark lines denote possible attachments of hard gluons, and the green ellipse represents $K\pi$-pair and the red one is the
light bachelor $D^0$ meson.}\label{fig:1}
\end{center}
\end{figure}

In light of the effective Hamiltonian eq.(\ref{hamiltionian}), we plot the Feynman diagrams contributing to quasi-two-body $B_{(s)}\to \overline{D}K\pi$ decays at the leading order in Figure~\ref{fig:1}. Based on the factorization formula and the wave functions of initial and final states, we then calculate the total decay amplitudes with the CKM matrix elements and the wilson coefficients, which are presented as
\begin{eqnarray}
\mathcal{A}(B^0\to\overline{D}^0(K^+\pi-))&=&\frac{G_F}{\sqrt{2}}V_{cb}^*V_{us}\left[(C_1+C_2/3)\mathcal{F}_{K\pi}^{LL}
+C_2\mathcal{M}_{K\pi}^{LL}\right],
\\
\mathcal{A}(B^0\to D^-(K^+\pi^0))&=&\frac{G_F}{\sqrt{2}}V_{cb}^*V_{us}\left[(C_1/3+C_2)\mathcal{F}_{D}^{LL}+C_1\mathcal{M}_{D}^{LL}\right],
\\
\mathcal{A}(B^0\to D_s^-(K^+\pi^0))&=&\frac{G_F}{\sqrt{2}}V_{cb}^*V_{ub}\left[(C_1+C_2/3)\mathcal{A}_{K\pi}^{LL}+C_2\mathcal{W}_{K\pi}^{LL}\right],
\\
\mathcal{A}(B^+\to\overline{D}^0(K^+\pi^0))&=&\frac{G_F}{\sqrt{2}}V_{cb}^*V_{us}\left[(C_1/3+C_2)\mathcal{F}_{D}^{LL}
+C_1\mathcal{M}_{D}^{LL}\right.\nonumber\\
&&\left.+(C_1+C_2/3)\mathcal{F}_{K\pi}^{LL}+C_2\mathcal{M}_{K\pi}^{LL}\right],
\\
\mathcal{A}(B_s\to \overline{D}^0(K^-\pi^+))&=&\frac{G_F}{\sqrt{2}}V_{cb}^*V_{ud}\left[(C_1+C_2/3)\mathcal{F}_{K\pi}^{LL}+C_2\mathcal{M}_{K\pi}^{LL}\right],
\\
\mathcal{A}(B_s\to D_s^-(K^+\pi^0))&=&\frac{G_F}{\sqrt{2}}V_{cb}^*V_{us}\left[(C_1/3+C_2)\mathcal{F}_{D}^{LL}+C_1\mathcal{M}_{D}^{LL}\right.\\
&&\left.+(C_1+C_2/3)\mathcal{A}_{K\pi}^{LL}+C_2\mathcal{W}_{K\pi}^{LL}\right].
\end{eqnarray}
In above decay amplitudes, the symbols $\mathcal{F}_{K\pi(D)}^{LL}$, $\mathcal{M}_{K\pi(D)}^{LL}$, $\mathcal{A}_{K\pi(D)}^{LL}$ and $\mathcal{W}_{K\pi(D)}^{LL}$ denote the contributions from the factorizable emission type diagrams, the hard-rescattering emission diagrams, the $W$ boson annihilation type diagrams, and the $W$ boson exchange type diagrams, respectively. The subscript $K\pi(D)$ represents the $K\pi$ pair ($D$ meson) is recoiled, and the superscript $LL$  stands for the contributions coming from $(V-A)\bigotimes(V-A)$ current. For the sake of simplicity, the expressions of the $\mathcal{F}_{K\pi(D)}^{LL}$, $\mathcal{M}_{K\pi(D)}^{LL}$, $\mathcal{A}_{K\pi(D)}^{LL}$ and $\mathcal{W}_{K\pi(D)}^{LL}$ are not given any more, which can be obtained by making the following substitutions in the results of ref.~\cite{Zou:2020atb},
\begin{eqnarray}
\phi_K^{a,p}\to\phi_D,\;\phi_K^t=0,\; r_K\to r_D=m_d/m_B.
\end{eqnarray}
We also point out that when considering the resonant contribution from the $D$-wave resonance $K_2^*(1430)$, the contribution $\mathcal{F}_{D}^{LL}$ disappears due to the fact that the tensor structure can not be produced through $(V-A)$ current.

\section{Numerical Results and Discussions}\label{sec:result}
In this section, we firstly list the parameters we used in the numerical calculations, including the QCD scale, the masses and the lifetimes of the $B/B_s$ mesons, the mass of the $\overline{D}$ meson, the masses and the widths of the intermediate resonances, and the CKM matrix elements,
\begin{eqnarray}
&&\Lambda_{QCD}^{f=4}=0.25\pm0.05~{\rm GeV},\;
m_{B}=5.279~ {\rm GeV},\;m_{B_s}=5.366~ {\rm GeV},\nonumber\\
&&\tau_{B^+}=1.638~ {\rm ps},\;\tau_{B^0}=1.519~ {\rm ps},\;\tau_{B_s}=1.520~ {\rm ps},\nonumber\\
&&m_{D^0}=1.865~ {\rm GeV},\;m_{D^+}=1.869~ {\rm GeV},\;m_{D_s}=1.96~ {\rm GeV},\nonumber\\
&&m_{K^*(892)}=0.892~ {\rm GeV},\;m_{K^*(1410)}=1.414~ {\rm GeV},\;m_{K_0^*(1430)}=1.425~ {\rm GeV},\nonumber\\
&&m_{K^*_2(1430)}=1.427~ {\rm GeV},\;\Gamma_{K^*(892)}=51.4~ {\rm MeV},\;\Gamma_{K^*(1410)}=232~ {\rm MeV},\nonumber\\
&&\Gamma_{K^*_0(1430)}=270~{\rm MeV},\;\Gamma_{K_2^*(1430)}=100~ {\rm MeV},\nonumber\\
&&V_{cb}=0.04182^{+0.00085}_{-0.00074},\;V_{ud}=0.97435\pm0.00016,\;V_{us}=0.22500\pm0.00067.
\end{eqnarray}

In Table.\ref{Table:br}, we present the numerical results of quasi-two-body $B_{(s)}\to \overline{D} K\pi$ decays, together with the experimental data from LHCb \cite{LHCb:2014ioa}. In our calculations, we take three kinds of errors into accounts to evaluate the theoretical uncertainties. The first uncertainties arise from the nonperturbative parameters in the wave functions of the initial and final states, such as the decay constants $f_B=(0.19\pm0.02)~{\rm GeV}$ and $f_{B_s}=(0.23\pm0.02)~{\rm GeV}$ of $B$ and $B_s$ meson respectively; the shape parameter $\omega_B=(0.4\pm0.04)~{\rm GeV}$ and $\omega_{B_s}=(0.5\pm0.05)~{\rm GeV}$ for $B$ meson and $B_s$ meson; the Gegenbauer moments in LCDAs of $K\pi$ pair and the $\overline{D}$ meson. We emphasize that this kind of uncertainties is dominant, but these uncertainties could  be reduced with the improvements of the experimental measurements and the developments of the nonperturbative approaches. The second uncertainties origin from the unknown contributions from QCD radiative corrections \cite{Li:2012nk,Li:2010nn} and the power corrections \cite{Shen:2018abs,Shen:2019vdc}, which are characterized by varying the $\Lambda_{\rm QCD}=0.25\pm0.05 ~{\rm GeV}$ and the factorization scale $t$ from $0.8t$ to $1.2t$. In recent years, there are some attempts to study these two kinds of corrections in two-body $B$ decays \cite{Cheng:2020fcx,Lu:2022fgz,Chai:2022kmq}, but the relate researches on three-body $B$ decays have not been carried out yet. The last uncertainties are caused by the uncertainties of the CKM matrix elements $V_{cb}$ and $V_{ud(s)}$, which have marginal effects on the branching fractions.

\begin{table}
\caption{The flavour-averaged branching ratios (in $10^{-5}$) of $B_{(s)}\to \overline{D}K\pi$ decays with resonances $K^*(892)/(1410)$, $K_0^*(1430)$ and $K_2^*(1430)$ in PQCD approach, together with the experimental data \cite{LHCb:2014ioa} from LHCb.} \label{Table:br}
\begin{center}
\begin{tabular}{l c c}
 \hline \hline
Decay Modes &PQCD  & EXP \\
\hline\hline
 $B^0 \to \overline{D}^0(K^{*0}(892)\to) K^+\pi^-$
 &$1.71^{+1.00+0.33+0.02}_{-0.82-0.31-0.08}$
 &....\\
 $B^0 \to \overline{D}^0(K^{*0}(1410)\to) K^+\pi^-$
 &$0.10^{+0.05+0.04+0.01}_{-0.04-0.03-0.01}$
 &....\\
 $B^0 \to \overline{D}^0K^+\pi^-(LASS)$
 &$2.01^{+1.10+0.10+0.20}_{-0.86-0.44-0.02}$
 & $.....$ \\
 $B^0 \to \overline{D}^0(K_0^{*0}(1430)\to) K^+\pi^-$
 &$1.58^{+0.81+0.27+0.15}_{-0.68-0.43-0.03}$
 & $....$  \\
 $B^0 \to \overline{D}^0K^+\pi^- (LASS NR)$
 & $1.10^{+0.60+0.20+0.12}_{-0.47-0.26-0.00}$
 &$.....$  \\
 $B^0 \to \overline{D}^0(K_2^{*0}(1430)\to)K^+ \pi^-$
 &$0.25^{+0.15+0.06+0.03}_{-0.10-0.05-0.00}$
 &$.....$   \\
\hline
 $B^0 \to D^-(K^{*+}(892)\to) K^+ \pi^0$
 &$12.4^{+5.5+1.5+0.6}_{-4.0-1.8-0.5}$
 &....\\
 $B^0 \to D^-(K^{*+}(1410)\to) K^+ \pi^0$
 &$0.80^{+0.35+0.09+0.03}_{-0.25-1.10-0.03}$
 &....\\
 $B^0 \to D^-K^+\pi^0(LASS)$
 &$0.18^{+0.08+0.00+0.01}_{-0.06-0.01-0.00}$
 & $.....$ \\
 $B^0 \to D^-(K_0^{*+}(1430)\to) K^+ \pi^0$
 &$0.12^{+0.06+0.02+0.01}_{-0.05-0.02-0.00}$
 & $....$  \\
 $B^0 \to D^-K^+\pi^0(LASS NR)$
 & $0.10^{+0.05+0.01+0.00}_{-0.04-0.01-0.00}$
 &$....$  \\
 $B^0 \to D^-(K_2^{*+}(1430)\to)K^+\pi^0$
 &$0.17^{+0.10+0.09+0.01}_{-0.08-0.08-0.00}$
 &$...$   \\
 \hline
 $B^0 \to D_s^-(K^{*+}(892)\to) K^+ \pi^0$
 &$5.15^{+1.41+0.46+0.45}_{-1.19-0.76-0.15}$
 &....\\
 $B^0 \to D_s^-(K^{*+}(1410)\to) K^+ \pi^0$
 &$0.36^{+0.10+0.04+0.02}_{-0.07-0.04-0.01}$
 &....\\
 $B^0 \to D_s^-K^+\pi^0(LASS)$
 &$0.23^{+0.11+0.03+0.02}_{-0.16-0.13-0.00}$
 & $.....$ \\
 $B^0 \to D_s^-(K_0^{*+}(1430)\to) K^+ \pi^0$
 &$0.21^{+0.10+0.03+0.01}_{-0.13-0.00-0.00}$
 & $....$  \\
 $B^0 \to D_s^-K^+\pi^0(LASS NR)$
 & $0.12^{+0.06+0.01+0.01}_{-0.07-0.01-0.00}$
 &$....$  \\
 $B^0 \to D_s^-(K_2^{*+}(1430)\to)K^+\pi^0$
 &$0.93^{+0.52+0.09+0.10}_{-0.36-0.06-0.00}$
 &$...$   \\
 \hline
 $B^+ \to \overline{D}^{0}(K^{*+}(892)\to) K^+ \pi^0$
 &$16.7^{+7.1+3.2+0.7}_{-5.3-3.4-0.7}$
 &....\\
 $B^+ \to \overline{D}^{0}(K^{*+}(1410)\to) K^+ \pi^0$
 &$1.23^{+0.51+0.24+0.05}_{-0.40-0.25-0.05}$
 &....\\
 $B^+ \to \overline{D}^{0}K^+\pi^0(LASS)$
 &$2.14^{+1.76+0.06+0.03}_{-1.08-0.27-0.00}$
 & $.....$ \\
 $B^+ \to \overline{D}^{0}(K_0^{*+}(1430)\to) K^+ \pi^0$
 &$1.03^{+0.61+0.19+0.11}_{-0.46-0.18-0.00}$
 & $....$  \\
 $B^+ \to \overline{D}^{0}K^+\pi^0(LASS NR)$
 & $1.20^{+0.90+0.16+0.12}_{-0.63-0.20-0.00}$
 &$....$  \\
 $B^+ \to \overline{D}^{0}(K_2^{*+}(1430)\to)K^+\pi^0$
 &$0.25^{+0.16+0.11+0.02}_{-0.10-0.08-0.00}$
 &$...$   \\
 \hline
 $B_s \to \overline{D}^{0}(\overline{K}^{*0}(892)\to) K^- \pi^+$
 &$28.6^{+16.7+4.3+0.5}_{-13.3-5.6-0.8}$
 &$28.6\pm0.6\pm0.7\pm0.9\pm4.2$\\
 $B_s \to \overline{D}^{0}(\overline{K}^{*0}(1410)\to) K^- \pi^+$
 &$1.74^{+0.97+0.54+0.03}_{-0.79-0.50-0.07}$
 &$1.7\pm0.5\pm0.2\pm1.4\pm0.2$\\
 $B_s \to \overline{D}^{0} K^-\pi^+(LASS)$
 &$25.7^{+16.1+3.0+2.9}_{-11.1-6.1-0.0}$
 & $21.4\pm1.4\pm1.0\pm4.7\pm3.1$ \\
 $B_s \to \overline{D}^{0}(\overline{K}_0^{*0}(1430)\to) K^- \pi^+$
 &$19.7^{+11.0+3.7+0.22}_{-8.5-6.9-0.00}$
 &$20.0\pm1.6\pm0.7\pm3.3\pm2.9$  \\
 $B_s \to \overline{D}^{0}K^-\pi^+(LASS NR)$
 &$14.1^{+8.7+2.8+1.7}_{-6.2-3.5-0.0}$
 &$13.7\pm2.5\pm1.5\pm4.1\pm2.0$  \\
 $B_s \to \overline{D}^{0}(\overline{K}_2^{*0}(1430)\to)K^-\pi^+$
 &$3.95^{+2.41+1.13+0.46}_{-1.65-0.96-0.00}$
 &$3.7\pm0.6\pm0.4\pm1.1\pm0.5$   \\
 \hline
 $B_s \to D_s^-(K^{*+}(892)\to) K^+ \pi^0$
 &$11.1^{+4.5+2.0+0.5}_{-3.3-2.1-0.4}$
 &....\\
 $B_s \to D_s^-(K^{*+}(1410)\to) K^+ \pi^0$
 &$0.66^{+0.27+0.13+0.03}_{-0.20-0.13-0.03}$
 &....\\
 $B_s \to D_s^-K^+\pi^0(LASS)$
 &$0.16^{+0.11+0.02+0.04}_{-0.05-0.03-0.00}$
 & $.....$ \\
 $B_s \to D^-(K_0^{*+}(1430)\to) K^+ \pi^0$
 &$0.12^{+0.08+0.03+0.02}_{-0.04-0.02-0.00}$
 & $....$  \\
 $B_s \to D^-K^+\pi^0(LASS NR)$
 &$0.09^{+0.05+0.02+0.02}_{-0.03-0.01-0.00}$
 &$....$  \\
 $B_s \to D_s^-(K_2^{*+}(1430)\to)K^+\pi^0$
 &$0.24^{+0.16+0.13+0.02}_{-0.10-0.08-0.00}$
 &$...$   \\
 \hline \hline
\end{tabular}
\end{center}
\end{table}

From Table.\ref{Table:br}, it can be seen that for the $B_s\to \overline{D}^0 (K^*\to) K^- \pi^+$ decays the branching fractions we calculated are in good agreement with the LHCb experimental measurements that are based on a data sample corresponding to an integrated luminosity of 3.0 $fb^{-1}$ of $pp$ collision data with respect to the $K^*(892)$, $K^*(1410)$, $K_0^*(1430)$ and $K^*_2(1430)$ resonances. Using the fitted wave functions of $K\pi$ pair we also calculate the branching fractions of other $B_{(s)}\to \overline{D} (K^*\to) K\pi$ decays, and they are expected to be tested in experiments in future. Due to the enhancement by the large CKM elements, most branching fractions are at the range $10^{-6}-10^{-4}$, which are measurable in LHCb and Belle-II experiments.

For the quasi-two-body $B_s\to \overline{D}^0(R\to) K^- \pi^+$ decays corresponding to the $S$, $P$ and $D$ wave intermediate resonances $K^*$, the branching fractions are larger than the other corresponding decays, because the quasi-two-body $B_s\to \overline{D}^0 K^- \pi^+$ decays are governed by the process $\bar{b} \to \bar{c} u \bar{d}$, which is enhanced by $\mid\frac{V_{ud}}{V_{us}}\mid^2$ compared to the other decays such as quasi-two-body $B^0\to \overline{D}^0K^+\pi^-$ decays governed by $\bar{b} \to \bar{c} u s$. As a result, the $B_s\to \overline{D}^0K^-\pi^+$ decays are firstly measured in the LHCb experiment. However, for the quasi-two-body $B^0 \to D_s^- K^+\pi^0 $ decays that are also governed by the $\bar{b} \to \bar{c} u d$ transition and enhanced by the large CKM elements $V_{cb}V_{ud}$, the branching fractions are far smaller than those of $B_s\to \overline{D}^0K^-\pi^+$ decays, because they are pure annihilation type decays. In particular, they are viewed as to be power suppressed in comparison with the emission diagrams. For the $B_s\to \overline{D}^0K^-\pi^+$ and $B^0\to \overline{D}^0 K^+\pi^-$ decays with the $\overline{D}$ emitted, the branching fractions corresponding to the $S$ and $D$ wave resonances are also sizable or even similar to that of the $P$ wave resonance. However, the $S$ and $D$ wave resonant branching fractions of the $B^0\to D^-K^+\pi^0$, $B^+\to \overline{D}^0 K^+\pi^0$ and $B_s\to D_s^-K^+\pi^0$ decays with $K\pi$ pair emitted are much smaller than that of $P$ wave resonant branching fractions, and it is  because the amplitudes with $S$ wave or $D$ wave $K-\pi$ pair emission are highly suppressed or forbidden.

In light of SU(3) symmetry, we define a ratio between the quasi-two-body $B_s\to \overline{D}^0(K^*\to)K^-\pi^+$ and $B^0\to \overline{D}^0 (K^*\to)K^+\pi^-$ decays, and the naive estimation is given as
\begin{eqnarray}
R=\frac{Br(B_s\to \overline{D}^0(K^*\to)K^-\pi^+)}{Br(B^0\to \overline{D}^0 (K^*\to)K^+\pi^-)}\sim|\frac{V_{ud}}{V_{us}}|^2\frac{\tau_{B_s}}{\tau_{B^0}}\sim18.
\end{eqnarray}
Accordingly, we also evaluate these ratios from Table.\ref{Table:br} as
\begin{eqnarray}
R=\frac{Br(B_s\to \overline{D}^0(\bar{K}^*(892)\to)K^-\pi^+)}{Br(B^0\to \overline{D}^0 (K^*(892)\to)K^+\pi^-)}\sim\frac{28.6^{+17.2}_{-14.4}}{1.71^{+1.05}_{-0.88}}\sim16.7^{+14.3}_{-12.1},
\label{eq:1}
\\
R=\frac{Br(B_s\to \overline{D}^0(\bar{K}^*(1410)\to)K^-\pi^+)}{Br(B^0\to \overline{D}^0 (K^*(1410)\to)K^+\pi^-)}\sim\frac{1.74^{+1.11}_{-0.93}}{0.10^{+0.06}_{-0.05}}\sim17.4^{+15.2}_{-12.7},
\\
R=\frac{Br(B_s\to \overline{D}^0(\bar{K}_0^*(1430)\to)K^-\pi^+)}{Br(B^0\to \overline{D}^0 (K_0^*(1430)\to)K^+\pi^-)}\sim\frac{19.7_{-10.9}^{+11.6}}{1.58_{-0.80}^{+0.87}}\sim12.5^{+10.0}_{-9.4},
\\
R=\frac{Br(B_s\to \overline{D}^0(\bar{K}_2^*(1430)\to)K^-\pi^+)}{Br(B^0\to \overline{D}^0 (K_2^*(1430)\to)K^+\pi^-)}\sim\frac{3.95_{-1.90}^{+2.70}}{0.25_{-0.11}^{+0.16}}\sim16.0^{+14.7}_{-10.3}.
\label{eq:4}
\end{eqnarray}
It is obvious that there are large uncertainties in the results eqs.~(\ref{eq:1} - \ref{eq:4}). The acceptable deviation between our calculations and naive estimations attributes to the complicate wave functions of the $K\pi$ pair. As aforementioned, although there are some attempts to study the wave functions of meson pair, the wave functions based on the first principle are still absent. Therefore, we have to employ  the phenomenological models that are introduced in Sec.\ref{sec:function}. In this regard, the accuracy of wave functions of heavy mesons should be further improved in future.

\begin{table}[!htb]
\caption{The branching ratios (in $10^{-5}$) of $B_{(s)}\to \overline{D}K^*$ decays probed from the quasi-two-body $B_{(s)}\to \overline{D}(K^*\to)K\pi$ decays based on the narrow-width-approximation(NWA),  together
with the experimental data \cite{Workman:2022ynf} and the former PQCD predictions from refs.\cite{Li:2008ts,Zou:2017iau,Zou:2012sx}} \label{br2}
\begin{center}
\begin{tabular}{l c c c}
 \hline \hline
 \multicolumn{1}{c}{Decay Modes}&\multicolumn{1}{c}{NWA} &\multicolumn{1}{c}{EXP} &\multicolumn{1}{c}{former PQCD} \\
\hline\hline
 $B^0 \to \overline{D}^0K^{*0}(892)$
 &$2.57^{+1.57}_{-1.32}$
 &$4.5\pm0.6$&$2.6_{-0.9}^{+1.1}$
 \\
 $B^0 \to \overline{D}^0K^{*0}(1410)$
 &$1.51^{+0.96}_{-0.81}$
 &$<6.7$&$...$
 \\
 $B^0 \to \overline{D}^0K_0^{*0}(1430)$
 &$2.55^{+1.41}_{-1.33}$
 &$0.7\pm0.7$&$3.19^{+3.21}_{-2.43}$
 \\
 $B^0 \to \overline{D}^0K_2^{*0}(1430)$
 &$0.75^{+0.49}_{-0.33}$   &$2.1\pm0.9$
 &$1.45_{-0.50}^{+0.51}$
 \\
\hline
 $B^0 \to D^-K^{*+}(892)$
 &$37.2^{+17.1}_{-13.2}$
 &$45\pm7$
 &$38.3^{+13.5}_{-13.9}$
 \\
 $B^0 \to D^-K^{*+}(1410)$
 &$36.3^{+17.8}_{-14.2}$
 &...
 &...
 \\
 $B^0 \to D^-K_0^{*+}(1430)$
 &$0.39^{+0.19}_{-0.17}$
 &$....$
 &$0.79_{-0.48}^{+0.40}$
 \\
 $B^0 \to D^-K_2^{*+}(1430)$
 &$1.02^{+0.78}_{-0.66}$
 &$...$
 &$1.16_{-0.62}^{+0.72}$
 \\
 \hline
 $B^0 \to D_s^-K^{*+}(892)$
 &$15.4^{+4.6}_{-4.2}$
 &$3.5\pm1.0$
 &$18.2^{+6.26}_{-6.45}$\\
 $B^0 \to D_s^-K^{*+}(1410)$
 &$16.3^{+5.9}_{-4.8}$
 &...
 &...
 \\
 $B^0 \to D_s^-K_0^{*+}(1430)$
 &$0.67^{+0.33}_{-0.42}$
 & $...$
 &$1.41^{+0.70}_{-0.65}$
 \\
 $B^0 \to D_s^-K_2^{*+}(1430)$
 &$5.59^{+3.18}_{-2.17}$
 &$...$
 &$6.06^{+1.81}_{-1.96}$
 \\
 \hline
 $B^+ \to \overline{D}^{0}K^{*+}(892)$
 &$50.1^{+23.4}_{-18.9}$
 &$53\pm4$
 &$63.7^{+20.5}_{-21.6}$
 \\
 $B^+ \to \overline{D}^{0}K^{*+}(1410)$
 &$55.9^{+27.4}_{-23.6}$
 &...
 &...
 \\
 $B^+ \to \overline{D}^{0}K_0^{*+}(1430)$
 &$3.32^{+2.14}_{-1.71}$
 & $...$
 &$4.72^{+2.46}_{-2.06}$
 \\
 $B^+ \to \overline{D}^{0}K_2^{*+}(1430)$
 &$1.50^{+1.14}_{-0.78}$
 &$...$
 &$3.33^{+1.60}_{-1.38}$
 \\
 \hline
 $B_s \to \overline{D}^{0}\overline{K}^{*0}(892)$
 &$42.9^{+25.8}_{-21.6}$
 &$44\pm6$
 &$43.6^{+22.0}_{-17.4}$
 \\
 $B_s \to \overline{D}^{0}\overline{K}^{*0}(1410)$
 &$39.5^{+26.4}_{-22.5}$
 &$39\pm35$
 &...
 \\
 $B_s \to \overline{D}^{0}\overline{K}_0^{*0}(1430)$
 &$31.9^{+19.0}_{-17.9}$
 & $30\pm7$&$53.9^{+27.2}_{-23.8}$
 \\
 $B_s \to \overline{D}^{0}\overline{K}_2^{*0}(1430)$
 &$11.8^{+8.1}_{-5.7}$
 &$11\pm4$
 &$20.3^{+8.7}_{-7.9}$
 \\
 \hline
 $B_s \to D_s^-K^{*+}(892)$
 &$33.3^{+14.7}_{-11.7}$
 &...
 &$28.1^{+16.7}_{-13.8}$\\

 $B_s \to D_s^-K^{*+}(1410)$
 &$30.0^{+14.8}_{-12.4}$
 &...
 &...
 \\
 $B_s \to D_s^-K_0^{*+}(1430)$
 &$0.38^{+0.25}_{-0.13}$
 & $....$
 &$0.39_{-0.16}^{+0.21}$
 \\
 $B_s \to D_s^-K_2^{*+}(1430)$
 &$1.44^{+1.26}_{-0.78}$
 &$...$
 &$1.97_{-0.96}^{+1.09}$
 \\
 \hline \hline
\end{tabular}
\end{center}
\end{table}

Under the narrow-width-approximation (NWA), the quasi-two-body decay and corresponding two-body process satisfy the factorization relation
\begin{eqnarray}
BF(B\to M_1R\to M_1M_2M_3)\approx BF(B\to M_1R)\times BF(R\to M_2M_3),
\end{eqnarray}
where the $R$ represents the resonance. Therefore, we could extract the branching fractions of the related two-body decays, by combining our numerical results and the corresponding branching fractions of the resonances decay to the $K\pi$. In turn, the comparison with the experimental data of two-body decays can help us to further verify the reliability of the wave functions of meson pair. In PDG \cite{Workman:2022ynf}, the branching fractions of the resonances decays to $K \pi$ are summarized as follows,
\begin{eqnarray}
BF(K^*(892)\to K\pi)&\sim&1,\\
BF(K^*(1410)\to K\pi)&=&(6.6\pm1.3)\%,\\
BF(K^*_0(1430)\to K\pi)&=&(93\pm10)\%,\\
BF(K_2^*(1430)\to K\pi)&=&(49.9\pm1.2)\%.
\end{eqnarray}
Using above data and the results in Table.\ref{Table:br}, we roughly extract the branching fractions of the corresponding two-body $B\to\overline{D}R$ decays and present the results in Table.~\ref{br2}. It can be seen that most branching fractions determined from NWA agree well with the available experimental data \cite{Workman:2022ynf} and the previous predictions \cite{Li:2008ts,Zou:2017iau,Zou:2012sx} based on PQCD within the uncertainties. Of course, if we wonder to test PQCD approach with these decays, the uncertainties in both theoretical and experimental sides should be reduced in future. We also note that for the pure annihilation $B^0\to D_s^-K^{*+}(892)$ decay channel, the branching fraction $(15.4^{+4.6}_{-4.2})\times 10^{-5}$ we obtained is much larger than the experimental data $(3.5\pm1.0)\times 10^{-5}$, though it is still in agreement with the previous study $(18.2^{+6.3}_{-6.4}) \times 10^{-5}$ \cite{Li:2008ts}. In fact, the $B^0\to D_s^-K^{*+}(892)$ decay occurs only with the annihilation diagrams, which are power suppressed. This deviation may indicate that the contributions from high power corrections become important, and we left this study to be our future work.

\section{Summary}\label{sec:summary}
In this work we have investigated the resonant contributions of the quasi-two-body $B\to D K\pi$ decays with the vector resonances $K^*(892)/(1410)$, the scalar resonance $K_0^*(1430)$ and the tensor resonance $K_2^*(1430)$. We first fitted the parameters in the wave functions of the $K\pi$ pair with the experimental results from LHCb. With the wave functions we then calculated the other quasi-two-body $B_{(s)}\to \overline{D}(K^*\to) K\pi$ decays. Most branching fractions are at the range of $10^{-6}\sim 10^{-4}$, which is very promising to be measured experimentally. With the narrow-width-approximation, we also explored the branching fractions of the corresponding two-body decays, and most results are in agreement with the available experimental data. We hope our results to be further tested in experiments, so as to test the wave functions of $K\pi$ pair and PQCD approach. Because all decays are governed by only tree operators, there are no  direct $CP$ asymmetries in these decays. If large $CP$ asymmetry is observed in experiments, it would be a signal of new physics beyond SM.
\section*{Acknowledgment}
This work is supported in part by the National Science Foundation of China under the Grant No.~11705159, and the Natural Science Foundation of Shandong province under the Grant No.~ZR2022MA035 and No.~ZR2019JQ04.

\bibliographystyle{bibstyle}
\bibliography{refs}

\providecommand{\href}[2]{#2}\begingroup\raggedright\begin{thebibliography}{10}

\bibitem{Grossman:2002aq}
Y.~Grossman, Z.~Ligeti, and A.~Soffer, {\it {Measuring $\gamma$ in $B^\pm \to
  K^\pm (KK^*)_D$ decays}},  {\em Phys. Rev. D} {\bf 67} (2003) 071301,
  [\href{https://arxiv.org/abs/hep-ph/0210433}{{\tt hep-ph/0210433}}].

\bibitem{BaBar:2009vfr}
{\bf BaBar} Collaboration, B.~Aubert et~al., {\it {Dalitz Plot Analysis of
  $B^\pm \to \pi^\pm\pi^\pm\pi^\mp$ Decays}},  {\em Phys. Rev. D} {\bf 79}
  (2009) 072006, [\href{https://arxiv.org/abs/0902.2051}{{\tt
  arXiv:0902.2051}}].

\bibitem{BaBar:2012iuj}
{\bf BaBar} Collaboration, J.~P. Lees et~al., {\it {Study of CP violation in
  Dalitz-plot analyses of $B^0 \to K^+K^-K^0_S$, $B^+\to K^+K^-K^+$, and
  $B^+\to K^0_S K^0_SK^+$}},  {\em Phys. Rev.} {\bf D85} (2012) 112010,
  [\href{https://arxiv.org/abs/1201.5897}{{\tt arXiv:1201.5897}}].

\bibitem{BaBar:2009jov}
{\bf BaBar} Collaboration, B.~Aubert et~al., {\it {Time-dependent amplitude
  analysis of $B^0 \to K^0_S \pi^+\pi^-$}},  {\em Phys. Rev.} {\bf D80} (2009)
  112001, [\href{https://arxiv.org/abs/0905.3615}{{\tt arXiv:0905.3615}}].

\bibitem{BaBar:2011vfx}
{\bf BaBar} Collaboration, J.~P. Lees et~al., {\it {Amplitude Analysis of
  $B^0\to K^+ \pi^- \pi^0$ and Evidence of Direct CP Violation in $B\to K^*
  \pi$ decays}},  {\em Phys. Rev. D} {\bf 83} (2011) 112010,
  [\href{https://arxiv.org/abs/1105.0125}{{\tt arXiv:1105.0125}}].

\bibitem{BaBar:2011ktx}
{\bf BaBar} Collaboration, J.~P. Lees et~al., {\it {Amplitude analysis and
  measurement of the time-dependent CP asymmetry of $B^0 \to K_S^0 K_S^0 K_S^0$
  decays}},  {\em Phys. Rev. D} {\bf 85} (2012) 054023,
  [\href{https://arxiv.org/abs/1111.3636}{{\tt arXiv:1111.3636}}].

\bibitem{BaBar:2008lpx}
{\bf BaBar} Collaboration, B.~Aubert et~al., {\it {Evidence for Direct CP
  Violation from Dalitz-plot analysis of $B^\pm \to K^\pm \pi^\mp \pi^\pm$}},
  {\em Phys. Rev. D} {\bf 78} (2008) 012004,
  [\href{https://arxiv.org/abs/0803.4451}{{\tt arXiv:0803.4451}}].

\bibitem{Belle:2010wis}
{\bf Belle} Collaboration, Y.~Nakahama et~al., {\it {Measurement of CP
  violating asymmetries in $B^0 \to K^+K^- K^0_S$ decays with a time-dependent
  Dalitz approach}},  {\em Phys. Rev. D} {\bf 82} (2010) 073011,
  [\href{https://arxiv.org/abs/1007.3848}{{\tt arXiv:1007.3848}}].

\bibitem{Belle:2008til}
{\bf Belle} Collaboration, J.~Dalseno et~al., {\it {Time-dependent Dalitz Plot
  Measurement of CP Parameters in $B^0 \to K^0_S \pi^+ \pi^-$ Decays}},  {\em
  Phys. Rev. D} {\bf 79} (2009) 072004,
  [\href{https://arxiv.org/abs/0811.3665}{{\tt arXiv:0811.3665}}].

\bibitem{Belle:2006ljg}
{\bf Belle} Collaboration, A.~Garmash et~al., {\it {Dalitz Analysis of
  Three-body Charmless $B^0 \to K^0 \pi^+ \pi^-$ Decay}},  {\em Phys. Rev. D}
  {\bf 75} (2007) 012006, [\href{https://arxiv.org/abs/hep-ex/0610081}{{\tt
  hep-ex/0610081}}].

\bibitem{Belle:2005rpz}
{\bf Belle} Collaboration, A.~Garmash et~al., {\it {Evidence for large direct
  CP violation in $B^\pm\to \rho(770)^0K^\pm$ from analysis of the three-body
  charmless $B^\pm\to K^\pm \pi^\pm\pi^\mp$ decay}},  {\em Phys. Rev. Lett.}
  {\bf 96} (2006) 251803, [\href{https://arxiv.org/abs/hep-ex/0512066}{{\tt
  hep-ex/0512066}}].

\bibitem{LHCb:2019sus}
{\bf LHCb} Collaboration, R.~Aaij et~al., {\it {Amplitude analysis of the $B^+
  \rightarrow \pi^+\pi^+\pi^-$ decay}},  {\em Phys. Rev. D} {\bf 101} (2020),
  no.~1 012006, [\href{https://arxiv.org/abs/1909.05212}{{\tt
  arXiv:1909.05212}}].

\bibitem{LHCb:2019jta}
{\bf LHCb} Collaboration, R.~Aaij et~al., {\it {Observation of Several Sources
  of $CP$ Violation in $B^+ \to \pi^+ \pi^+ \pi^-$ Decays}},  {\em Phys. Rev.
  Lett.} {\bf 124} (2020), no.~3 031801,
  [\href{https://arxiv.org/abs/1909.05211}{{\tt arXiv:1909.05211}}].

\bibitem{LHCb:2017hbp}
{\bf LHCb} Collaboration, R.~Aaij et~al., {\it {Resonances and $CP$ violation
  in $B_s^0$ and $\overline{B}_s^0 \to J/\psi K^+K^-$ decays in the mass region
  above the $\phi(1020)$}},  {\em JHEP} {\bf 08} (2017) 037,
  [\href{https://arxiv.org/abs/1704.08217}{{\tt arXiv:1704.08217}}].

\bibitem{LHCb:2019vww}
{\bf LHCb} Collaboration, R.~Aaij et~al., {\it {Amplitude analysis of
  $B^{0}_{s} \rightarrow K^{0}_{\textrm{S}} K^{\pm}\pi^{\mp}$ decays}},  {\em
  JHEP} {\bf 06} (2019) 114, [\href{https://arxiv.org/abs/1902.07955}{{\tt
  arXiv:1902.07955}}].

\bibitem{LHCb:2016vqn}
{\bf LHCb} Collaboration, R.~Aaij et~al., {\it {Observation of the decay $B^0_s
  \to \phi\pi^+\pi^-$ and evidence for $B^0 \to \phi\pi^+\pi^-$}},  {\em Phys.
  Rev. D} {\bf 95} (2017), no.~1 012006,
  [\href{https://arxiv.org/abs/1610.05187}{{\tt arXiv:1610.05187}}].

\bibitem{El-Bennich:2009gqk}
B.~El-Bennich, A.~Furman, R.~Kaminski, L.~Lesniak, B.~Loiseau, and
  B.~Moussallam, {\it {CP violation and kaon-pion interactions in $B \to K
  \pi^+ \pi^-$ decays}},  {\em Phys. Rev. D} {\bf 79} (2009) 094005,
  [\href{https://arxiv.org/abs/0902.3645}{{\tt arXiv:0902.3645}}]. [Erratum:
  Phys.Rev.D 83, 039903 (2011)].

\bibitem{Krankl:2015fha}
S.~Kr\"ankl, T.~Mannel, and J.~Virto, {\it {Three-body non-leptonic B decays
  and QCD factorization}},  {\em Nucl. Phys. B} {\bf 899} (2015) 247--264,
  [\href{https://arxiv.org/abs/1505.04111}{{\tt arXiv:1505.04111}}].

\bibitem{Cheng:2002qu}
H.-Y. Cheng and K.-C. Yang, {\it {Nonresonant three-body decays of D and B
  mesons}},  {\em Phys. Rev. D} {\bf 66} (2002) 054015,
  [\href{https://arxiv.org/abs/hep-ph/0205133}{{\tt hep-ph/0205133}}].

\bibitem{Cheng:2016shb}
H.-Y. Cheng, C.-K. Chua, and Z.-Q. Zhang, {\it {Direct CP Violation in
  Charmless Three-body Decays of $B$ Mesons}},  {\em Phys. Rev. D} {\bf 94}
  (2016), no.~9 094015, [\href{https://arxiv.org/abs/1607.08313}{{\tt
  arXiv:1607.08313}}].

\bibitem{Cheng:2014uga}
H.-Y. Cheng and C.-K. Chua, {\it {Charmless three-body decays of $B_s$
  mesons}},  {\em Phys. Rev. D} {\bf 89} (2014), no.~7 074025,
  [\href{https://arxiv.org/abs/1401.5514}{{\tt arXiv:1401.5514}}].

\bibitem{Li:2014oca}
Y.~Li, {\it {Comprehensive study of $\overline B^0\to K^0(\overline K^0)
  K^\mp\pi^\pm$ decays in the factorization approach}},  {\em Phys. Rev. D}
  {\bf 89} (2014), no.~9 094007, [\href{https://arxiv.org/abs/1402.6052}{{\tt
  arXiv:1402.6052}}].

\bibitem{Li:2014fla}
Y.~Li, {\it {Branching Fractions and Direct $CP$ Asymmetries of $\bar B_s ^0
  \to K^0 h^+h^{\prime -}(h^{(\prime)}=K,\pi)$ Decays}},  {\em Sci. China Phys.
  Mech. Astron.} {\bf 58} (2015), no.~3 031001,
  [\href{https://arxiv.org/abs/1401.5948}{{\tt arXiv:1401.5948}}].

\bibitem{Wang:2014ira}
W.-F. Wang, H.-C. Hu, H.-n. Li, and C.-D. L\"u, {\it {Direct CP asymmetries of
  three-body $B$ decays in perturbative QCD}},  {\em Phys. Rev. D} {\bf 89}
  (2014), no.~7 074031, [\href{https://arxiv.org/abs/1402.5280}{{\tt
  arXiv:1402.5280}}].

\bibitem{Wang:2016rlo}
W.-F. Wang and H.-n. Li, {\it {Quasi-two-body decays $B\to K\rho\to K\pi\pi$ in
  perturbative QCD approach}},  {\em Phys. Lett. B} {\bf 763} (2016) 29--39,
  [\href{https://arxiv.org/abs/1609.04614}{{\tt arXiv:1609.04614}}].

\bibitem{Li:2016tpn}
Y.~Li, A.-J. Ma, W.-F. Wang, and Z.-J. Xiao, {\it {Quasi-two-body decays
  $B_{(s)}\to P\rho\to P\pi\pi$ in perturbative QCD approach}},  {\em Phys.
  Rev. D} {\bf 95} (2017), no.~5 056008,
  [\href{https://arxiv.org/abs/1612.05934}{{\tt arXiv:1612.05934}}].

\bibitem{Rui:2017bgg}
Z.~Rui, Y.~Li, and W.-F. Wang, {\it {The S-wave resonance contributions in the
  $B^0_s$ decays into $ \psi(2S,3S)$ plus pion pair}},  {\em Eur. Phys. J. C}
  {\bf 77} (2017), no.~3 199, [\href{https://arxiv.org/abs/1701.02941}{{\tt
  arXiv:1701.02941}}].

\bibitem{Zou:2020atb}
Z.-T. Zou, Y.~Li, Q.-X. Li, and X.~Liu, {\it {Resonant contributions to
  three-body $B\rightarrow KKK$ decays in perturbative QCD approach}},  {\em
  Eur. Phys. J. C} {\bf 80} (2020), no.~5 394,
  [\href{https://arxiv.org/abs/2003.03754}{{\tt arXiv:2003.03754}}].

\bibitem{Zou:2020fax}
Z.-T. Zou, Y.~Li, and X.~Liu, {\it {Branching fractions and CP asymmetries of
  the quasi-two-body decays in $B_{s} \rightarrow K^0({\overline{K}}^0)K^\pm
  \pi ^\mp $ within PQCD approach}},  {\em Eur. Phys. J. C} {\bf 80} (2020),
  no.~6 517, [\href{https://arxiv.org/abs/2005.02097}{{\tt arXiv:2005.02097}}].

\bibitem{Zou:2020mul}
Z.-T. Zou, Y.~Li, and H.-n. Li, {\it {Is $f_X(1500)$ observed in the $B\to
  \pi(K)KK$ decays $\rho^0(1450)$?}},  {\em Phys. Rev. D} {\bf 103} (2021),
  no.~1 013005, [\href{https://arxiv.org/abs/2007.13141}{{\tt
  arXiv:2007.13141}}].

\bibitem{Zou:2020ool}
Z.-T. Zou, L.~Yang, Y.~Li, and X.~Liu, {\it {Study of Quasi-two-body
  $B_{(s)}\to \phi (f_0(980)/f_2(1270)\to)\pi\pi$ Decays in Perturbative QCD
  Approach}},  {\em Eur. Phys. J. C} {\bf 81} (2021), no.~1 91,
  [\href{https://arxiv.org/abs/2011.07676}{{\tt arXiv:2011.07676}}].

\bibitem{Yang:2021zcx}
L.~Yang, Z.-T. Zou, Y.~Li, X.~Liu, and C.-H. Li, {\it {Quasi-two-body
  $B_{(s)}\to V \pi\pi$ decays with resonance $f_0(980)$ in the PQCD
  approach}},  {\em Phys. Rev. D} {\bf 103} (2021), no.~11 113005,
  [\href{https://arxiv.org/abs/2103.15031}{{\tt arXiv:2103.15031}}].

\bibitem{Liu:2021sdw}
W.-F. Liu, Z.-T. Zou, and Y.~Li, {\it {Charmless Quasi-Two-Body B Decays in
  Perturbative QCD Approach: Taking $B\to K(R\to K+K^-)$ as Examples}},  {\em
  Adv. High Energy Phys.} {\bf 2022} (2022) 5287693,
  [\href{https://arxiv.org/abs/2112.00315}{{\tt arXiv:2112.00315}}].

\bibitem{Zhang:2013oqa}
Z.-H. Zhang, X.-H. Guo, and Y.-D. Yang, {\it {CP violation in $B^{\pm}
  \rightarrow \pi^{\pm}\pi^{+}\pi^{-}$ in the region with low invariant mass of
  one $\pi^{+}\pi^{-}$ pair}},  {\em Phys. Rev. D} {\bf 87} (2013), no.~7
  076007, [\href{https://arxiv.org/abs/1303.3676}{{\tt arXiv:1303.3676}}].

\bibitem{El-Bennich:2006rcn}
B.~El-Bennich, A.~Furman, R.~Kaminski, L.~Lesniak, and B.~Loiseau, {\it
  {Interference between $f_0(980)$ and $\rho(770)^0$ resonances in $B\to \pi^+
  \pi^- K$ decays}},  {\em Phys. Rev. D} {\bf 74} (2006) 114009,
  [\href{https://arxiv.org/abs/hep-ph/0608205}{{\tt hep-ph/0608205}}].

\bibitem{Cheng:2019tgh}
S.~Cheng and J.-M. Shen, {\it {$\bar{B}_s \rightarrow f_0(980)$ form factors
  and the width effect from light-cone sum rules}},  {\em Eur. Phys. J. C} {\bf
  80} (2020), no.~6 554, [\href{https://arxiv.org/abs/1907.08401}{{\tt
  arXiv:1907.08401}}].

\bibitem{Hu:2022eql}
R.~Hu and Z.-H. Zhang, {\it {Data-based analysis of the forward-backward
  asymmetry in $B^\pm\to K^\pm K^\mp K^\pm$}},  {\em Phys. Rev. D} {\bf 105}
  (2022), no.~9 093007, [\href{https://arxiv.org/abs/2201.07456}{{\tt
  arXiv:2201.07456}}].

\bibitem{Snyder:1993mx}
A.~E. Snyder and H.~R. Quinn, {\it {Measuring CP asymmetry in $B \to \rho \pi$
  decays without ambiguities}},  {\em Phys. Rev. D} {\bf 48} (1993) 2139--2144.

\bibitem{Cheng:2007si}
H.-Y. Cheng, C.-K. Chua, and A.~Soni, {\it {Charmless three-body decays of B
  mesons}},  {\em Phys. Rev. D} {\bf 76} (2007) 094006,
  [\href{https://arxiv.org/abs/0704.1049}{{\tt arXiv:0704.1049}}].

\bibitem{Cheng:2008vy}
H.-Y. Cheng, {\it {Theoretical Overview of Hadronic Three-body B Decays}},
  [\href{https://arxiv.org/abs/0806.2895}{{\tt arXiv:0806.2895}}].

\bibitem{Herndon:1973yn}
D.~Herndon, P.~Soding, and R.~J. Cashmore, {\it {A GENERALIZED ISOBAR MODEL
  FORMALISM}},  {\em Phys. Rev. D} {\bf 11} (1975) 3165.

\bibitem{Chung:1995dx}
S.~U. Chung, J.~Brose, R.~Hackmann, E.~Klempt, S.~Spanier, and C.~Strassburger,
  {\it {Partial wave analysis in K matrix formalism}},  {\em Annalen Phys.}
  {\bf 4} (1995) 404--430.

\bibitem{Belle:2004drb}
{\bf Belle} Collaboration, A.~Garmash et~al., {\it {Dalitz analysis of the
  three-body charmless decays $B^+ \to K^+ \pi^+ \pi^-$ and $B^+ \to K^+ K^+
  K^-$ }},  {\em Phys. Rev. D} {\bf 71} (2005) 092003,
  [\href{https://arxiv.org/abs/hep-ex/0412066}{{\tt hep-ex/0412066}}].

\bibitem{BaBar:2009pnd}
{\bf BaBar} Collaboration, B.~Aubert et~al., {\it {Dalitz Plot Analysis of
  $B^-\to D^+ \pi^- pi^-$}},  {\em Phys. Rev. D} {\bf 79} (2009) 112004,
  [\href{https://arxiv.org/abs/0901.1291}{{\tt arXiv:0901.1291}}].

\bibitem{LHCb:2014ioa}
{\bf LHCb} Collaboration, R.~Aaij et~al., {\it {Dalitz plot analysis of $B_s^0
  \rightarrow \bar{D}^0 K^- \pi^+$ decays}},  {\em Phys. Rev. D} {\bf 90}
  (2014), no.~7 072003, [\href{https://arxiv.org/abs/1407.7712}{{\tt
  arXiv:1407.7712}}].

\bibitem{LHCb:2018oeb}
{\bf LHCb} Collaboration, R.~Aaij et~al., {\it {Observation of the decay $B_s^0
  \to \overline{D}^0 K^+ K^-$}},  {\em Phys. Rev. D} {\bf 98} (2018), no.~7
  072006, [\href{https://arxiv.org/abs/1807.01891}{{\tt arXiv:1807.01891}}].

\bibitem{LHCb:2013svv}
{\bf LHCb} Collaboration, R.~Aaij et~al., {\it {Measurement of the branching
  fractions of the decays $B^0_s \to \overline{D}^0K^-\pi^+$ and $B^0 \to
  \overline{D}^0K^+\pi^-$}},  {\em Phys. Rev. D} {\bf 87} (2013), no.~11
  112009, [\href{https://arxiv.org/abs/1304.6317}{{\tt arXiv:1304.6317}}].

\bibitem{Keum:2000ms}
Y.-Y. Keum and H.-n. Li, {\it {Nonleptonic charmless $B$ decays: Factorization
  versus perturbative QCD}},  {\em Phys. Rev. D} {\bf 63} (2001) 074006,
  [\href{https://arxiv.org/abs/hep-ph/0006001}{{\tt hep-ph/0006001}}].

\bibitem{Keum:2000ph}
Y.-Y. Keum, H.-n. Li, and A.~I. Sanda, {\it {Fat penguins and imaginary
  penguins in perturbative QCD}},  {\em Phys. Lett. B} {\bf 504} (2001) 6--14,
  [\href{https://arxiv.org/abs/hep-ph/0004004}{{\tt hep-ph/0004004}}].

\bibitem{Lu:2000em}
C.-D. Lu, K.~Ukai, and M.-Z. Yang, {\it {Branching ratio and CP violation of $B
  \to \pi \pi$ decays in perturbative QCD approach}},  {\em Phys. Rev. D} {\bf
  63} (2001) 074009, [\href{https://arxiv.org/abs/hep-ph/0004213}{{\tt
  hep-ph/0004213}}].

\bibitem{Li:2003yj}
H.-n. Li, {\it {QCD aspects of exclusive B meson decays}},  {\em Prog. Part.
  Nucl. Phys.} {\bf 51} (2003) 85--171,
  [\href{https://arxiv.org/abs/hep-ph/0303116}{{\tt hep-ph/0303116}}].

\bibitem{Buchalla:1995vs}
G.~Buchalla, A.~J. Buras, and M.~E. Lautenbacher, {\it {Weak decays beyond
  leading logarithms}},  {\em Rev. Mod. Phys.} {\bf 68} (1996) 1125--1144,
  [\href{https://arxiv.org/abs/hep-ph/9512380}{{\tt hep-ph/9512380}}].

\bibitem{Yu:2005rh}
X.-Q. Yu, Y.~Li, and C.-D. Lu, {\it {Branching ratio and CP violation of $B_s
  \to \pi K$ decays in the perturbative QCD approach}},  {\em Phys. Rev. D}
  {\bf 71} (2005) 074026, [\href{https://arxiv.org/abs/hep-ph/0501152}{{\tt
  hep-ph/0501152}}]. [Erratum: Phys.Rev.D 72, 119903 (2005)].

\bibitem{Ali:2007ff}
A.~Ali, G.~Kramer, Y.~Li, C.-D. Lu, Y.-L. Shen, W.~Wang, and Y.-M. Wang, {\it
  {Charmless non-leptonic $B_s$ decays to $PP$, $PV$ and $VV$ final states in
  the pQCD approach}},  {\em Phys. Rev. D} {\bf 76} (2007) 074018,
  [\href{https://arxiv.org/abs/hep-ph/0703162}{{\tt hep-ph/0703162}}].

\bibitem{Zou:2015iwa}
Z.-T. Zou, A.~Ali, C.-D. Lu, X.~Liu, and Y.~Li, {\it {Improved Estimates of The
  $B_{(s)}\to V V$ Decays in Perturbative QCD Approach}},  {\em Phys. Rev. D}
  {\bf 91} (2015) 054033, [\href{https://arxiv.org/abs/1501.00784}{{\tt
  arXiv:1501.00784}}].

\bibitem{Zou:2009zza}
H.~Zou, R.-H. Li, X.-X. Wang, and C.-D. Lu, {\it {The CKM suppressed $B(B_s)
  \to \overline D_{(s)}P, \overline D_{(s)}V, \overline D_{(s)}^*P, \overline
  D_{(s)}^*V$ decays in perturbative QCD approach}},  {\em J. Phys. G} {\bf 37}
  (2010) 015002, [\href{https://arxiv.org/abs/0908.1856}{{\tt
  arXiv:0908.1856}}].

\bibitem{Li:2008ts}
R.-H. Li, C.-D. Lu, and H.~Zou, {\it {The $B(B_s)) \to D_{(s)}P, D_{(s)}V,
  D_{(s)}^*P$ and $D_{(s)}^*V$ decays in the perturbative QCD approach}},  {\em
  Phys. Rev. D} {\bf 78} (2008) 014018,
  [\href{https://arxiv.org/abs/0803.1073}{{\tt arXiv:0803.1073}}].

\bibitem{Zou:2017yxc}
Z.-T. Zou, Y.~Li, and X.~Liu, {\it {Study of $B_c \to DS$ decays in the
  perturbative QCD approach}},  {\em Phys. Rev. D} {\bf 97} (2018), no.~5
  053005, [\href{https://arxiv.org/abs/1712.02239}{{\tt arXiv:1712.02239}}].

\bibitem{Zou:2017iau}
Z.-T. Zou, Y.~Li, and X.~Liu, {\it {Cabibbo-Kobayashi-Maskawa-favored $B$
  decays to a scalar meson and a $D$ meson}},  {\em Eur. Phys. J. C} {\bf 77}
  (2017), no.~12 870, [\href{https://arxiv.org/abs/1704.03967}{{\tt
  arXiv:1704.03967}}].

\bibitem{Zou:2016yhb}
Z.-T. Zou, Y.~Li, and X.~liu, {\it {Two-body charmed $B_{(s)}$ decays involving
  a light scalar meson}},  {\em Phys. Rev. D} {\bf 95} (2017), no.~1 016011,
  [\href{https://arxiv.org/abs/1609.06444}{{\tt arXiv:1609.06444}}].

\bibitem{Zou:2012sy}
Z.-T. Zou, X.~Yu, and C.-D. Lu, {\it {The $B_c\to D^{(*)}T$ decays in
  perturbative QCD approach}},  {\em Phys. Rev. D} {\bf 87} (2013) 074027,
  [\href{https://arxiv.org/abs/1208.4252}{{\tt arXiv:1208.4252}}].

\bibitem{Zou:2012sx}
Z.-T. Zou, X.~Yu, and C.-D. Lu, {\it {The $B(B_{s})\rightarrow
  D_{(s)}(\bar{D}_{(s)}) T$ and $D_{(s)}^{*}(\bar{D}_{(s)}^{*})T$ Decays in
  Perturbative QCD Approach}},  {\em Phys. Rev. D} {\bf 86} (2012) 094001,
  [\href{https://arxiv.org/abs/1205.2971}{{\tt arXiv:1205.2971}}].

\bibitem{Zou:2012td}
Z.-T. Zou, X.~Yu, and C.-D. Lu, {\it {Nonleptonic two-body charmless B decays
  involving a tensor meson in the Perturbative QCD Approach}},  {\em Phys. Rev.
  D} {\bf 86} (2012) 094015, [\href{https://arxiv.org/abs/1203.4120}{{\tt
  arXiv:1203.4120}}].

\bibitem{Liu:2013cvx}
X.~Liu, Z.-J. Xiao, and Z.-T. Zou, {\it {Branching ratios and CP violations of
  $B\to K^*_0(1430)K^*$ decays in the perturbative QCD approach}},  {\em Phys.
  Rev. D} {\bf 88} (2013), no.~9 094003,
  [\href{https://arxiv.org/abs/1309.7256}{{\tt arXiv:1309.7256}}].

\bibitem{Wang:2017hxe}
C.~Wang, Q.-A. Zhang, Y.~Li, and C.-D. Lu, {\it {Charmless $B_{(s)}\to VV$
  Decays in Factorization-Assisted Topological-Amplitude Approach}},  {\em Eur.
  Phys. J. C} {\bf 77} (2017), no.~5 333,
  [\href{https://arxiv.org/abs/1701.01300}{{\tt arXiv:1701.01300}}].

\bibitem{Li:2019hnt}
Y.~Li, D.-C. Yan, Z.~Rui, and Z.-J. Xiao, {\it {$S$, $P$ and $D$-wave resonance
  contributions to $B_{(s)} \to \eta_c(1S,2S) K\pi$ decays in the perturbative
  QCD approach}},  {\em Phys. Rev. D} {\bf 101} (2020), no.~1 016015,
  [\href{https://arxiv.org/abs/1911.09348}{{\tt arXiv:1911.09348}}].

\bibitem{Li:2021cnd}
Y.~Li, D.-C. Yan, J.~Hua, Z.~Rui, and H.-n. Li, {\it {Global determination of
  two-meson distribution amplitudes from three-body B decays in the
  perturbative QCD approach}},  {\em Phys. Rev. D} {\bf 104} (2021), no.~9
  096014, [\href{https://arxiv.org/abs/2105.03899}{{\tt arXiv:2105.03899}}].

\bibitem{Wang:2020saq}
W.-F. Wang, J.~Chai, and A.-J. Ma, {\it {Contributions of $K^*_0(1430)$ and
  $K^*_0(1950)$ in the three-body decays $B\to K\pi h$}},  {\em JHEP} {\bf 03}
  (2020) 162, [\href{https://arxiv.org/abs/2001.00355}{{\tt
  arXiv:2001.00355}}].

\bibitem{Meadows:2007jm}
B.~Meadows, {\it {Low Mass S-wave $K \pi$ and $\pi \pi$ System}},  {\em eConf}
  {\bf C070805} (2007) 27, [\href{https://arxiv.org/abs/0712.1605}{{\tt
  arXiv:0712.1605}}].

\bibitem{Aston:1987ir}
D.~Aston et~al., {\it {A Study of $K^- \pi^+$ Scattering in the Reaction $K^- p
  \to K^- \pi^+ n$ at 11-GeV/c}},  {\em Nucl. Phys. B} {\bf 296} (1988)
  493--526.

\bibitem{Back:2017zqt}
J.~Back et~al., {\it {LAURA$^{++}$: A Dalitz plot fitter}},  {\em Comput. Phys.
  Commun.} {\bf 231} (2018) 198--242,
  [\href{https://arxiv.org/abs/1711.09854}{{\tt arXiv:1711.09854}}].

\bibitem{112}
W.~Dunwoodie, {\it {Fits to $K\pi$ $I= 1/2$ S-wave amplitude and phase data,
  http://www.slac.stanford.edu/wmd/kpiswave/ kpiswavefit.note}}, .

\bibitem{Blatt:1952ije}
J.~M. Blatt and V.~F. Weisskopf, {\em {Theoretical nuclear physics}}.
\newblock Springer, New York, 1952.

\bibitem{BaBar:2005qms}
{\bf BaBar} Collaboration, B.~Aubert et~al., {\it {Dalitz-plot analysis of the
  decays $B^\pm \to K^\pm \pi^\mp \pi^\pm$}},  {\em Phys. Rev. D} {\bf 72}
  (2005) 072003, [\href{https://arxiv.org/abs/hep-ex/0507004}{{\tt
  hep-ex/0507004}}]. [Erratum: Phys.Rev.D 74, 099903 (2006)].

\bibitem{Cheng:2010hn}
H.-Y. Cheng, Y.~Koike, and K.-C. Yang, {\it {Two-parton Light-cone Distribution
  Amplitudes of Tensor Mesons}},  {\em Phys. Rev. D} {\bf 82} (2010) 054019,
  [\href{https://arxiv.org/abs/1007.3541}{{\tt arXiv:1007.3541}}].

\bibitem{Li:2012nk}
H.-n. Li, Y.-L. Shen, and Y.-M. Wang, {\it {Next-to-leading-order corrections
  to $B \to \pi$ form factors in $k_T$ factorization}},  {\em Phys. Rev. D}
  {\bf 85} (2012) 074004, [\href{https://arxiv.org/abs/1201.5066}{{\tt
  arXiv:1201.5066}}].

\bibitem{Li:2010nn}
H.-n. Li, Y.-L. Shen, Y.-M. Wang, and H.~Zou, {\it {Next-to-leading-order
  correction to pion form factor in $k_T$ factorization}},  {\em Phys. Rev. D}
  {\bf 83} (2011) 054029, [\href{https://arxiv.org/abs/1012.4098}{{\tt
  arXiv:1012.4098}}].

\bibitem{Shen:2018abs}
Y.-L. Shen, Z.-T. Zou, and Y.-B. Wei, {\it {Subleading power corrections to
  $B\to \gamma l\nu $ decay in PQCD approach}},  {\em Phys. Rev. D} {\bf 99}
  (2019), no.~1 016004, [\href{https://arxiv.org/abs/1811.08250}{{\tt
  arXiv:1811.08250}}].

\bibitem{Shen:2019vdc}
Y.-L. Shen, Z.-T. Zou, and Y.~Li, {\it {Power Corrections to Pion Transition
  Form Factor in Perturbative QCD Approach}},  {\em Phys. Rev. D} {\bf 100}
  (2019), no.~1 016022, [\href{https://arxiv.org/abs/1901.05244}{{\tt
  arXiv:1901.05244}}].

\bibitem{Cheng:2020fcx}
S.~Cheng and Z.-J. Xiao, {\it {The PQCD approach towards to next-to-leading
  order: A short review}},  {\em Front. Phys. (Beijing)} {\bf 16} (2021), no.~2
  24201, [\href{https://arxiv.org/abs/2009.02872}{{\tt arXiv:2009.02872}}].

\bibitem{Lu:2022fgz}
C.-D. L\"u, Y.-L. Shen, C.~Wang, and Y.-M. Wang, {\it {Enhanced
  Next-to-Leading-Order Corrections to Weak Annihilation $B$-Meson Decays}},
  [\href{https://arxiv.org/abs/2202.08073}{{\tt arXiv:2202.08073}}].

\bibitem{Chai:2022kmq}
J.~Chai, S.~Cheng, Y.-h. Ju, D.-C. Yan, C.-D. L\"u, and Z.-J. Xiao, {\it
  {Charmless two-body $B$ meson decays in perturbative QCD factorization
  approach}},  [\href{https://arxiv.org/abs/2207.04190}{{\tt
  arXiv:2207.04190}}].

\bibitem{Workman:2022ynf}
{\bf Particle Data Group} Collaboration, R.~L. Workman, {\it {Review of
  Particle Physics}},  {\em PTEP} {\bf 2022} (2022) 083C01.

\end{thebibliography}\endgroup
\end{document}